# Film Coating Process Research and Characterization of TiN Coated Racetrack-type Ceramic Pipe*

WANG Jie（王洁）, XU Yan-Hui（徐延辉）, ZHANG Bo（张波）, WEI Wei（尉伟）, FAN Le（范乐）, PEI Xiang-Tao（裴香涛）, HONG Yuan-Zhi（洪远志）, WANG Yong（王勇）[1]

National Synchrotron Radiation Laboratory, University of Science and Technology of China, HeFei, AnHui 230029 China

* Work supported by the National Nature Science Foundation of China under Grant Nos.11075157.

1) corresponding author: ywang@ustc.edu.cn

**ABSTRACT**

TiN film was coated on the internal face of racetrack-type ceramic pipe by three different methods: radio-frequency sputtering, DC sputtering and DC magnetron sputtering. The deposition rates of TiN film under different coating methods were compared. According to the AFM, SEM, XPS test results, these properties were analyzed, such as TiN film roughness and surface morphology. At the same time, the deposition rates were studied under two types' cathode, Ti wires and Ti plate. According to the SEM test results, Ti plate cathode can improve the TiN/Ti film deposition rate obviously.

## INTRODUCTION

In the past few decades, TiN thin film caused great interest because of its low secondary electron yield (SEY), good electrical conductivity, stability of performance, ability to block hydrogen permeation, etc[1-3] The properties and coating process of TiN thin film have been studied in KEK [1], NSRL [4], DESY [5], BNL [6], etc. TiN coated ceramic vacuum chamber is the key equipment of the electron storage ring injection system at Hefei Light Source II (HLS II). However, for some irregular type ducts, such as the racetrack type (Fig. 1) ceramic chamber in accelerators, the shape of the ceramic pipeline will induce considerable new technological difficulties for the uniformity of TiN coating which is important for the vacuum and beam stability in the pipeline. Therefore, it has an extremely valuable research that how to get the film which has high quality and meets the requirements of the physical design of the storage ring, such as mitigating the electron cloud instability [7].

In the present research, TiN film was deposited on ceramic pipe with different cathode structures and various film coating methods. What is more, the TiN film properties were investigated by atomic force microscope (AFM), scanning electron microscope (SEM) and X-ray photoelectron spectroscopy (XPS).

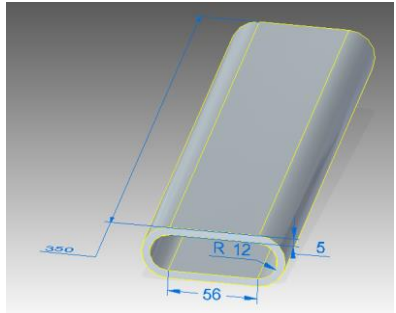

Figure1: A diagram of a ceramic vacuum pipeline.

## APPARATUS AND METHODS

### Coating system description

TiN films were deposited onto the interior wall of ceramic vacuum pipe which is described in Fig. 1. The deposition system which is shown in Fig. 2 consists of observation window, 300 l/s turbo molecular pump, power supply, vacuum gauge, coating vacuum chamber and gas flow control system. Argon gas and nitrogen are introduced into the sputtering system through an adjustable leak valve. There are two types of cathode structure, three Ti wires and one Ti plate whose size is 2 * 25 * 56mm$^3$. Figure 3 shows the position of silicon samples on the ceramic pipe inner surface. In addition, the silicon samples were used to do SEM tests conveniently.

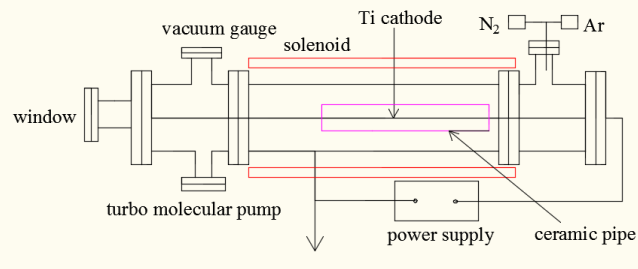

Figure 2: Schematic diagram of sputtering coating system.

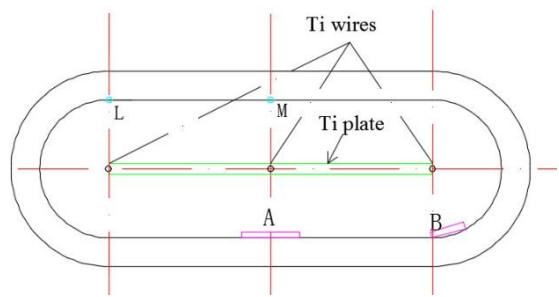

Figure3: The position of Silicon samples on the ceramic pipe inner surface, three Ti wires cathode and Ti plate cathode.

### Characterization Method

Thickness were measured by use of a Sirion 200 Schottky field scanning electron microscope (SEM). SEM basic parameters used in the experiments are as follows: 5kV accelerating voltage, 5.4mm working distance, 91.53KX magnification. Besides, material surface and internal compositional data were taken with a Thermo ESCALAB 250 X-ray photoelectron spectroscopy (XPS). The spectrometer was equipped with a hemispherical analyzer, a monochromater, a beam spot size of 500 $\mu m$ and all XPS data was measured with Al Ka X-rays with ($h\nu$ =1486.6 eV) operated at 150 W and an analyzer at 45 degrees. Surface morphology was observed through Innova atomic force microscope (AFM) at room temperature. The micrographs were taken at 5 kV to keep the images surface sensitive.

## Cathode structure

The cross-section of shaped ceramic vacuum chambers in the system is a special type of racetrack with a small aperture. In order to obtain a uniform thin film, a few titanium wires which are horizontally mounted as cathode target in the experiment have been used. Because of the non-rotational symmetry, the coated film could not be uniform obviously with one or two titanium wires as cathodes. So three cathodes have been considered. Due to the complexity of the calculation process, the analysis of sputtering rates(S) on different locations of the inner surface of ceramic vacuum chamber have been achieved by using the Matlab software.

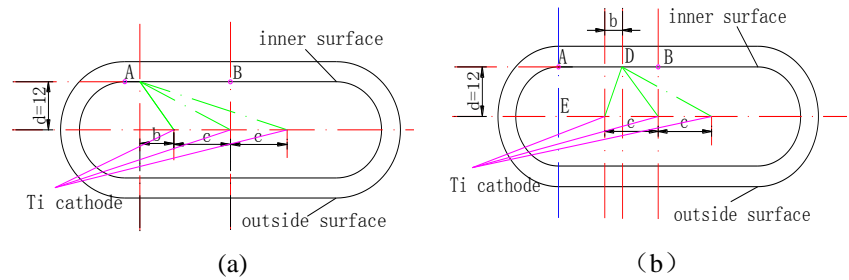

(a)　　　　　　　　　　　　　(b)

Figure 4: The sputtering rate analysis of one point on the inner surface of a ceramic vacuum pipeline, which c is the distance between the Ti cathodes, b is the distance of horizontal direction between the left Ti cathode and the point on the inner surface. (a)The point is on the left of the left Ti cathode in the horizontal direction.　(b) The point is on the right of the left Ti cathode in the horizontal direction.

The left side of ceramic vacuum pipeline have been simulated according to the characteristics of symmetry. In case (a), the larger b, S is smaller in point B. In case (b), $0 \leq b \leq c$, in point D, $S=S_1+S_2+S_3$

$$S_1 = \frac{\pi K d}{2(b^2+d^2)}, S_2 = \frac{\pi K d}{2\left[(c-b)^2+d^2\right]}, S_3 = \frac{\pi K d}{2\left[(2c-b)^2+d^2\right]}$$

where $K$ is constant. $S$ take a great value when b=c for point B and $S$ increase with c for point A. Above all, point E is an appropriate location for left Ti cathodes considering the homogeneity of the film thickness. On base of computation, the ratio of the maximum and the minimum film thickness is about 1.4:1 for c=28. Meanwhile the average thickness of the TiN film in point A and B is 20.6 nm 27.9 nm in dc sputtering experiments, which means that the average thickness ratio is 1:1.35. Hence the results are in good agreement with the experimental data.

　　　In order to improve the efficiency of sputtering, titanium plate cathode has been used. Furthermore, CST software are adopted to simulate the distribution of the electric field under different size of titanium plate. According to the simulation result, titanium plate size with 66 * 2mm can meet the requirements of the film uniformity, because the electric potential was basically the same on the edge of the ceramic (Figure 5).

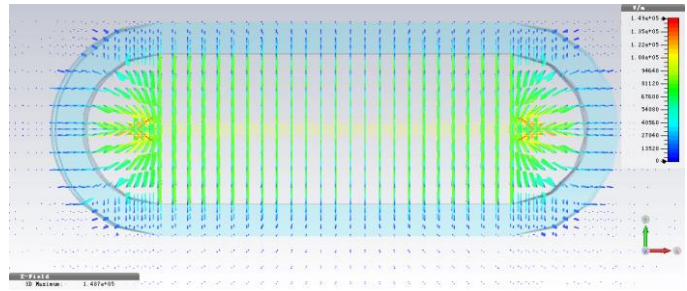

Figure 5: Electric field distribution in the case of Ti plate with a size of 66*2*350 mm$^3$.

## RESULTS AND DISCUSSION

### DC sputtering

During the experiment, the background vacuum pressure was below $10^{-4}$ Pa, then mixed gases of nitrogen and argon were injected which were controlled by D08-3B/ZM gas mass flow. The typical sputtering parameters are: -900 V cathode voltage, 66 Pa gas pressure, 3h deposition time, 5:1 nitrogen and argon gas flow ratio and 30 mA sputtering current. The film thickness was tested by selecting two points each sample at position A and B which were showed in Figure 6. The average thickness of the film was 51.0 nm. The approximate deposition rate is about 17 nm/h. Figure 6 shows the cross-sectional SEM morphology of TiN film sample prepared by DC sputtering.

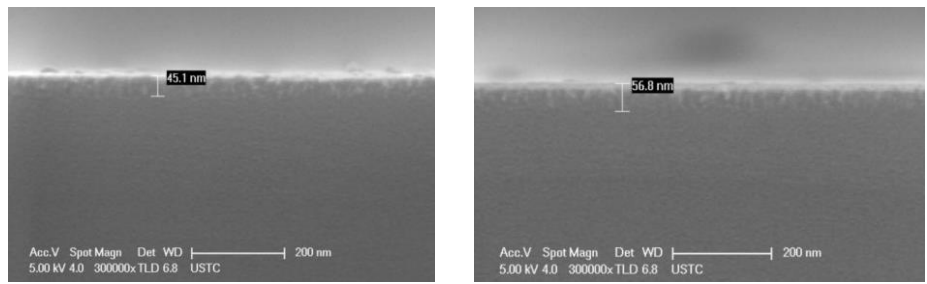

Figure 6: The cross-sectional SEM morphology of TiN film sample prepared by DC sputtering with three Ti wires cathode.

### Radio-frequency sputtering

TiN film thickness was 750 ~ 800 nm which can be seen in figure 7 and deposition rate was about 80 nm/h by radio-frequency sputtering.

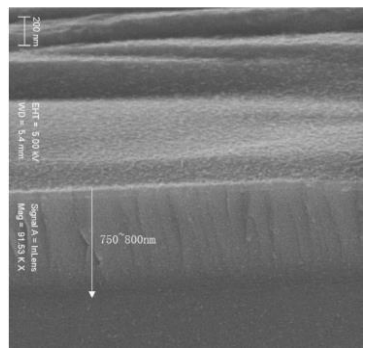

Figure 7: The cross-sectional SEM morphology of TiN film sample prepared by radio-frequency sputtering with three Ti wires cathode, 10h deposition time.

## DC magnetron sputtering

The typical sputtering parameters are: -542 ~ -503 V cathode voltage, 20~30 Pa working pressure, 3h deposition time, 2.5 Sccm nitrogen and 5 Sccm argon, 200 Gauss magnetic field and 0.5 A sputtering current. The film deposition rate and thickness with different cathode type was shown in table 1. Figure 8 shows the cross-sectional SEM morphology of TiN film sample prepared by DC magnetron sputtering with Ti plate cathode. In figure 9, the roughness of TiN films on position A and B which were prepared by DC magnetron sputtering with Ti plate cathode, were 157.7 nm and 233.3 nm within 5 $\mu m$ range respectively. Film deposition rate can be improved significantly using Ti plate cathode by DC magnetron sputtering which was shown in table 1.

Table 1: Under various deposition methods, the film deposition rate and thickness were compared with different cathode type.

| deposition method | deposition rate (nm/h) | thickness on position A/nm | thickness on position B/nm | cathode type |
|---|---|---|---|---|
| Radio-frequency sputtering | 80 | 800 | 750 | three Ti wires |
| DC sputtering | 17 | 56.8 | 45.1 | three Ti wires |
| DC magnetron sputtering | 156 | 780 | 535 | three Ti wires |
| DC magnetron sputtering | 800 | 3200 | 2720 | Ti plate |

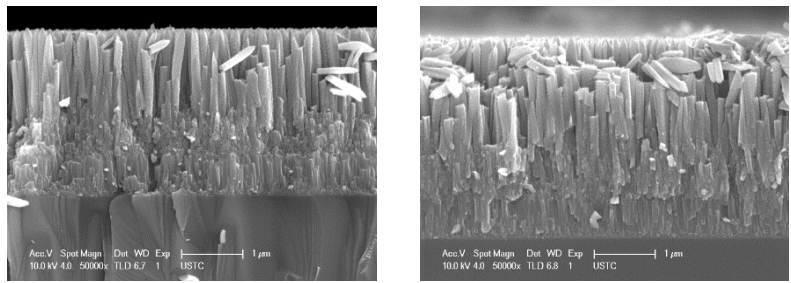

Figure 8: The cross-sectional SEM morphology of TiN films on position A (left) and B (right) prepared by DC magnetron sputtering with Ti plate cathode.

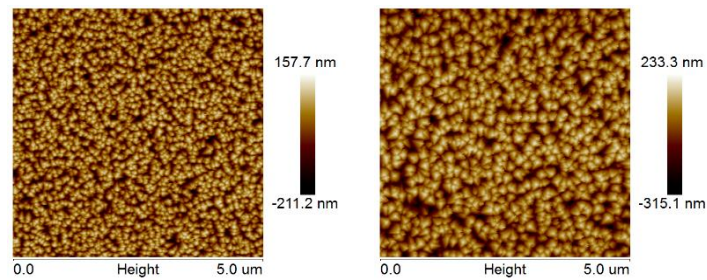

Figure 9: The AFM morphology of TiN films on position A (left) and B (right) prepared by DC magnetron sputtering with Ti plate cathode.

## CONCLUSION

For non-conductive ceramic material, TiN film deposition rate can be effectively improved by DC magnetron sputtering method，compared to dc sputtering and RF sputtering method. In addition, cathode installation position, the shape and size of the cathode material, can influence the uniformity of film thickness, to a certain extent. TiN film deposition rate by DC sputtering was about 20 nm/h with three Ti wires cathode, while it was 156 nm/h by DC magnetron sputtering. Besides, DC magnetron sputtering combined with Ti plate cathode, can greatly improve the film deposition rate which was 800 nm/h on the basis of cross-sectional SEM morphology. AFM test results shows that the roughness of TiN was about 200 nm within 5 $\mu m$ range which meet the requirements of engineering in accelerator field [8].